\documentclass[pra,aps,preprint,showpacs,articles]{revtex4}
\usepackage{epsf,epsfig}
\usepackage[psamsfonts]{amssymb}
\usepackage{amsmath}

\newcommand{\ket}[1]{|#1\rangle}
\newcommand{\bra}[1]{\langle #1|}
\newcommand{\proj}[2]{\ket{#1}\bra{#2}}

\newcommand{\outprod}[2]{\ket{#1}\bra{#2}}

\newcommand{\Rtil}{\tilde{\rho}}

\begin{document}

\title{NON-EQUILIBRIUM THERMAL ENTANGLEMENT DYNAMICS OF HEISENBERG SYSTEMS}
\author{Fardin
Kheirandish\footnote{fardin$_{-}$kh@phys.ui.ac.ir}, S. Javad
Akhtarshenas \footnote{akhtarshenas@phys.ui.ac.ir} and Hamidreza
Mohammadi \footnote{h.mohammadi@sci.ui.ac.ir}
}\affiliation{Department of Physics, University of Isfahan,
 Hezar Jarib Ave., Isfahan, Iran}
\begin{abstract}
\noindent The effects of initial conditions and system parameters on
entanglement dynamics and asymptotic entanglement for a two-qubit
anisotropic XY Heisenberg system in the presence of an inhomogeneous
magnetic field and spin-orbit interaction are investigated. We
suppose that each qubit interacts with a separate thermal reservoir
which is held in its own temperature. The effects of the parameters
of the system and environment like spin-orbit interaction and
temperature difference of reservoirs are also discussed.
\end{abstract}

 \maketitle

\section{INTRODUCTION}

Entanglement is a central theme in quantum information processing
which is  first noted by Schr\"{o}dinger \cite{EPR, S}. It
strongly affects our conceptual implication on physics, and force
us to change significantly our perspective of Nature. Entanglement
implies that the best knowledge of the whole of a composite system
may not include complete knowledge of its parts. In mathematical
sense a pure state of pair of quantum systems is called entangled
if it is unfactorizable. A mixed state $\rho$ of a bipartite
system is said to be separable or classically correlated if it can
be expressed as a convex combination of uncorrelated states
$\rho_A$ and $\rho_B$ of each subsystems i.e. $\rho=\sum_{i}
\,\omega_i \rho_A ^i \otimes \rho_B ^i$ such that $\omega_i \geq
0$ and $\sum_{i}\,\omega_i =1$, otherwise $\rho $ is entangled
\cite {NC-book,A-book}. Entanglement has no classical analog and
then can be considered as a uniquely quantum mechanical
(non-classical) resource that plays a key role in many of the most
interesting applications of quantum computation and quantum
information processing such as: quantum teleportation,
entanglement teleportation, quantum cryptography, and etc.
\cite{NC-book,A-book}. Performance the above mentioned tasks needs
to quantifying and optimizing the amounts of the entanglements in
a suitable multipartite quantum system. Many measures of
entanglement have been introduced and analyzed \cite
{NC-book,W,E-thesis}, but the one most relevant to this work is
entanglement of formation, which is intended to quantify the
resources need to create a given entangled state \cite {W}. For
the case of a two-qubit system Wootters \cite {W} has shown that
entanglement of formation can be obtained explicitly as: $E(\rho )
= \Xi [C(\rho )] = h\left( {\frac{{1 + \sqrt {1 + C^2 } }}{2}}
\right), $ where $h(x)=-x\log_2 x - (1-x)\log_2 (1-x)$ is the
binary entropy function and
$C(\rho)=\max\{0,2\lambda_{max}-\sum_{i=1}^{4}\,\lambda_i\}$ is
the concurrence of the state, where $\lambda_i$s are positive
square roots of the eigenvalues of the non-Hermitian matrix
$R=\rho \Rtil$, and $\tilde{\rho}$ is defined by $
\Rtil:=(\sigma^y \otimes \sigma^y)\rho^* (\sigma^y \otimes
\sigma^y)$. The function $\Xi$ is a monotonically increasing
function and ranges from 0 to 1 as C goes from 0 to 1, so that one
can take the concurrence as a measure of entanglement in its own
right. In the case that the state of the system is pure i.e.
$\rho=\proj {\psi}{\psi}$, $\ket{\psi}=a \ket{00}+b \ket{01}+c
\ket {10}+d \ket{11}$, the above formula is simplified to
$C(\ket{\psi})=2
\mid ad-bc \mid$. \\
Real Quantum systems are not isolated from their environment.
Unavoidable interaction between system and its environment cause to
leakage the coherence of the system to environment and hence
quantum-to-classical transition (see \cite{M-book} and references
therein) . More precisely, this interaction will, in general, create
some entanglement between the states of quantum system and the huge
states of environment and as a consequence, quantum coherence
initially localized within the system will become a shared property
of composite system-environment state and can no longer be observed
at the level of the system, leading to decoherence. Decoherence
destroys the quantumness of the system and hence will decrease the
useful entanglement between the parts of the system. A given
dynamics for a composite quantum system can exhibit several distinct
properties for asymptotic entanglement behavior, like: asymptotic
death of entanglement(ADE) i.e. entanglement vanishes exponentially
in time, entanglement sudden death (ESD) i.e. entanglement vanishes
faster than local coherence of the system, asymptotically steady
state entanglement(ASE) i.e the entanglement of the system reaches a
stationary value, asymptotically, entanglement sudden birth (ESB)
i.e. an initially separable state acquire entanglement
asymptotically and etc \cite{DT, T}.
\\
Among the numerous concepts considered to implement quantum bits
(qubits), approaches based on semiconductor quantum dots (QDs) offer
the great advantage that ultimately a miniaturized version of a
quantum computer is feasible. Indeed, Loss and DiViencenzo initially
proposed a quantum computer protocol based on electron spins trapped
in semiconductor QDs in 1998 \cite{LDi, DD, CL}. Here qubit is
represented by the spin of a single electron in a QD, which can be
initialized, manipulated, and read out by extremely sensitive
devices. Such systems are more scalable and have a longer coherence
time than other systems such as quantum optical and NMR systems. In
this paper quantum entanglement dynamics of an open two-qubit system
is realized by considering two electrons confined in two coupled
quantum dots (CQDs) interacting with two separate reservoirs. We
refer to this two-qubit system as "nanosystem" in the rest of the
paper. Because of weak lateral confinement electrons can tunnel from
one dot to the other and spin-spin and spin-orbit interactions
between the two qubits exist. Thus we can model this nanosystem by a
two-qubit spin chain including spin-orbit interaction. The spin
chain is the natural candidates for the realization of entanglement
and Heisenberg model is the simplest method for studying and
investigating the behavior of the spin chains. On the other hand,
 in what follows we model the environment by a thermal
reservoir and  assume that the inter dot separation is large enough
such that each  dot couples to a separate thermal reservoir (bosonic
bath). Here the bathes are assumed to be in thermodynamical
equilibrium at different temperature $\beta_i=\frac{1}{k_B T_i}$. In
general, there are two different ways for connecting the quantum
dots to the reservoirs: (i) "\textit{direct geometry}"; where a high
temperature bath is in contact with a QD in the presence of a strong
magnetic field i.e. $b \Delta T>0$ and (ii) "\textit{indirect
geometry}"; where a high temperature bath is in contact with a QD in
the presence of a weak magnetic field i.e. $b \Delta T<0$. The
 results show that the non-equilibrium  thermal
entanglement dynamics depends on the geometry of the system. \\
Entanglement properties of Heisenberg systems at thermal equilibrium
(thermal entanglement) are extensively studied after Nielsen
\cite{N}, who first studied the entanglement of a two-qubit
Heisenberg XXX chain (see \cite{Hamid} and references therein) . For
non-equilibrium thermal entanglement in spin systems, Eisler et al.
\cite{Ei} calculated the von Neumann entropy of a block of spins in
XX spin chain in the presence of the energy current and showed that
the enhancement of the amount of entanglement due to an energy
current is possible. After them, the non-equilibrium thermal
entanglement for steady state of some systems has been studied in a
number of works \cite{WBPS,SMM,P}. For example, Quiroga in Ref.
\cite{QR} considered a simple spin chain system (XXX-Heisenberg)
which is in contact with two different heat reservoirs and showed
that for the steady state, a temperature gradient can  increase or
decrease entanglement depending on the internal coupling strength
between spins. Dynamics of non-equilibrium thermal entanglement of
the same system has been studied by Sinaysky et al.\cite{Si}. They
have derived an analytical expression for the density matrix of the
system at a finite time. They also have shown that the system
converges to a steady state, asymptotically and the amount of
entanglement of the steady state takes its maximum value for unequal
bath temperatures and also the local energy levels can maintain the
entanglement at higher temperatures. However, the non-equilibrium
thermal entanglement dynamics of more involved spin systems (e,g XY
and XYZ Heisenberg systems) has not been considered, yet. In this
paper, we will investigate dynamics of non-equilibrium thermal
entanglement of a two qubit anisotropic XY Heisenberg system
(nanosystem) in the presence of the inhomogeneous magnetic field and
the spin orbit interaction. The influence of the parameters of the
nanosystem (i.e. magnetic field (B), inhomogeneity of magnetic
field(b), partial anisotropy($\chi$), mean coupling (J) and the
spin-orbit interaction parameter (D)) and environmental parameters
(i.e. temperatures $T_1$ and $T_2$ or equally $T_M$ and $\Delta T$,
and the couplings $\gamma_1$ and $\gamma_2$) on the entanglement of
the nanosystem is investigated. We have shown that, there is a
steady state entanglement for asymptotically large times. The size
of this steady state (asymptotic) entanglement and the dynamical
behavior of entanglement depend on the parameters of the model and
also on the geometry of the system. Increasing the temperature
difference $\Delta T$, and mean temperature $T_M$, decrease the
amount of asymptotic entanglement. We have also shown that, the size
of $T_M^{cr.}$ (temperature over which the entanglement vanishes)
and the amount of entanglement can be improved by adjusting the
value of the spin-orbit interaction parameter $D$. The maximum
entanglement ($C=1$) can be achieved for the case of large values of
$D$ and zero temperature reservoirs($T_1=T_2=0$). Furthermore, we
find that the indirect geometry is more suitable for creating and
maintaining the entanglement. The results obtained here are
consistent with those obtained in \cite{Hamid, QR, Si}.
\\
The paper is organized as follows. In Sec. II we introduce the
Hamiltonian of the whole system-reservoir under the rotating wave
approximation and then write the Markovian master equation governed
on the nanosystem by tracing out the reservoirs' degrees of freedom.
Ultimately, given some initial states, the density matrix of the
nanosystem at a later time is derived exactly. The effects of
initial conditions and system parameters on the dynamics of
entanglement and entanglement of asymptotic state of the nanosystem
are presented in Sec. III. Finally in Sec. IV a discussion concludes
the paper.

\section{THE MODEL AND THE HAMILTONIAN}

The total Hamiltonian of the nanosystem  which is interacting with
two heat reservoirs (bosonic bath) is described by
\begin{eqnarray}\label{Hamiltonian 0}
 \hat{H} =\hat{H}_S +\hat{H}_{B1}+ \hat{H}_{B2}+\hat{H}_{SB1}+\hat{H}_{SB2},
 \end{eqnarray}
where $\hat H_S$ is the Hamiltonian of the nanosystem, $\hat H_{Bj}$
is the Hamiltonian of the jth bath (j=1,2) and $\hat H_{SBj}$
denotes the system-bath interaction Hamiltonian. Nanosystem consists
of two spin electron confined in a two coupled quantum dots and is
described by a two-qubit anisotropic Heisenberg XY-model in the
presence of an inhomogeneous magnetic field and spin-orbit
interaction \cite{Hamid} with the following Hamiltonian
\begin{eqnarray}\label{Hamiltonian 1}
 \hat{H} _S&=& {\textstyle{1 \over 2}}(J_x \,\sigma _1^x \sigma _2^x \,
 + J_y \,\sigma _1^y \sigma _2^y +\textbf{B} _1 \cdot \boldsymbol {\sigma}
 _1+
  \textbf{B} _2 \cdot \boldsymbol {\sigma} _2
 \nonumber\\&+& \textbf{D} \cdot (\boldsymbol {\sigma} _1  \times
  \boldsymbol{\sigma}  _2
 )+ \delta \,\, \boldsymbol {\sigma}_1 \cdot \overline{\mathbf{\Gamma}}\cdot \boldsymbol
 {\sigma}_2)
 \end{eqnarray}
where $\boldsymbol{\sigma}_{j}=(\sigma^{x}_{j}, \sigma^{y}_{j},
\sigma^{z}_{j})$ is the vector of Pauli matrices, $\textbf{B}_j
\,(j=1,2)$  is the magnetic field on site j, $J_\mu \,(\mu=x,y)$ are
the real coupling coefficients (the chain is anti-ferromagnetic
(AFM) for $J_\mu >0$ and ferromagnetic (FM) for $J_\mu <0$)  and
$\textbf{D}$ is  Dzyaloshinski-Moriya vector, which is of first
order in spin-orbit coupling and is proportional to the coupling
coefficients ($J_\mu$) and $\overline{\mathbf{\Gamma}}$ is a
symmetric tensor which is of second order in spin-orbit coupling
\cite{D,M1,M2,M3}. For simplicity we assume $\textbf{B}_j =B_j \,
\boldsymbol{\hat{z}}$ such that $B_1 =B+b$ and $B_2 =B-b$, where b
indicates the amount of inhomogeneity of magnetic field. The vector
$\textbf{D}$ and the parameter $\delta$ are dimensionless, in system
like coupled GaAs quantum dots
$\boldsymbol{|}\textbf{D}\boldsymbol{|}$ is of order of a few
percent, while the order of last term is $10^{-4}$ and is
negligible. If we take $\textbf{D}= J D \, \boldsymbol{\hat{z}}$ and
ignore the second order spin-orbit coupling, then the above
Hamiltonian can be written as:

\begin{eqnarray}\label{Hamiltonian 2}
 \hat{H}_S &=& J\chi (\sigma _1^ +  \sigma _2^ +   + \sigma_1^ -  \sigma _2^ -  )
 + J(1 + i D)\sigma _1^ +\sigma _2^ -   + J(1 - i D)\sigma _1^ -  \sigma _2^ +
  \nonumber\\&+& (\frac{{B + b}}{2})\sigma _1^z
  + (\frac{{B - b}}{2})\sigma _2^z,
 \end{eqnarray}
where $J :=\frac{J_x + J_y}{2}$, is the mean coupling coefficient in
the XY-plane, $\chi :=\frac{J_x - J_y}{J_x + J_y}$, specifies the
amount of anisotropy in the XY-plane (partial anisotropy, $-1\leq
\chi \leq 1$) and $\sigma^\pm=\frac{1}{2}(\sigma^x \pm i\sigma^y)$
are lowering and raising operators.
 The spectrum of $H_S$ is easily obtained as
\begin{eqnarray}\label{spectrum}
\begin{array}{l}
\ket{\varepsilon _{1,2}}= \ket{\Psi ^ \pm}  = N^ \pm  ( (\frac{{b
\pm \xi }}{{J(1 - i D)}})\ket{01}  +
\ket{10})\,,\,\,\,\,\,\,\,\varepsilon  _{1,2}  =   \pm \xi\,,
 \\ \\
\ket{\varepsilon _{3,4}}= \ket{\Sigma ^ \pm}  = M^ \pm  (
(\frac{{B \pm \eta }} {{J\chi }})\ket{00}  + \ket{11}
)\,,\,\,\,\,\,\,\,\,\,\,\,\,\,\, \varepsilon  _{3,4}  =  \pm
\eta\,.
 \\
 \end{array}
\end{eqnarray}
Here the eigenstates are expressed in the standard basis
$\{\ket{00},\ket{01},\ket{10},\ket{11}\}$. In the above equations
$N^\pm =(1+\frac{{(b \pm \xi)^2}}{J^2 +(J D)^2})^{-1/2}$ and
$M^\pm =(1+(\frac{{B \pm \eta}}{J \chi})^2)^{-1/2}$ are the
normalization constants. Here we define, $ \xi := (b^2 + J^2 + (J
D)^2 )^{1/2}$ and
$\eta :=(B^2  + (J\chi )^2 )^{1/2}$, for later convenience.\\
The Hamiltonian of the reservoirs for each spin $j=1,2$ are given by
\begin{eqnarray}\label{Hamiltonian 3}
 \hat{H}_{Bj} =\sum_n \omega_n \hat{b}_{nj}^\dag \hat{b}_{nj}.
 \end{eqnarray}
 The interaction between the nanosystem and the bosonic
  bathes in the rotating wave approximation is
as following
\begin{eqnarray}\label{Hamiltonian 4}
 \hat{H}_{SBj} =\sigma_j^+ \sum_n g_n^{(j)}\hat{b}_{n,\,j}+\sigma_j^- \sum_n
 g_n^{(j)*}\hat{b}_{n,\,j}^\dag\equiv \sum_\mu \hat{V}_{j,\,\mu}\textit{\^{f}}_{j,\,\mu}
\end{eqnarray}
The system operator $\hat{V}_{j,\,\mu}$ are chosen to satisfy
$[\hat{H}_S,\hat{V}_{j,\,\mu}]=\omega_{j,\,\mu} \hat{V}_{j,\,\mu}$,
and the $\textit{\^{f}}_{j,\,\mu}$s are the random operators of
reservoirs and act on the bath degrees of freedom. Physically, the
index $\mu$ corresponds to transitions between the internal levels
of the nanosystem induced by the bath. The dynamics of the whole
nanosystem+reservoirs is described by a density operator
($\hat{\sigma}$) satisfying the Liouville equation $\dot{\hat
\sigma}=-i[\hat{H},\hat{\sigma}]$. If the coupling strengths of the
nanosystem and the environment are weak, the evolution of the
nanosystem does not influence the states of the reservoirs and one
can write $\hat{\sigma}(t)=\hat{\rho}(t)\hat{\rho}_{B1}(0)
\hat{\rho}_{B2}(0)$ (irreversibility hypothesis), where
$\hat{\rho}(t)$ is the reduced density matrix describing the
nanosystem and each bosonic bath is described by a canonical density
matrix of the form $\hat{\rho}_{Bj} = \emph{e}^{-\beta_j
\hat{H}_{Bj}}/Z$, where $Z=Tr(\emph{e}^{-\beta_j
\hat{H}_{Bj}})$ is the partition function of the jth bath.\\
In the Born-Markov approximation the master equation describing
the dynamics of the reduced density matrix of the nanosystem is
\cite{BP-book,Gold}
\begin{eqnarray}\label{master equation}
\frac{d\hat{\rho}}{dt}=-i[\hat{H}_S,\hat{\rho}]+\pounds_1(\hat{\rho})+\pounds_2(\hat{\rho})
\end{eqnarray}
where $\pounds_j(\hat{\rho})$ are  \textit{dissipators} given by
\cite{Gold}

\begin{eqnarray}\label{dissipiators 1}
\pounds_j(\hat{\rho})\equiv\sum_{\mu,\,\nu}
J_{\mu,\,\nu}^{(j)}(\omega_{j,\,\nu})
\{[\hat{V}_{j,\,\mu},[\hat{V}_{j,\,\nu}^\dag,\hat{\rho}]]-(1-\emph{e}^{\beta_j
\omega_{j,\,\nu}})[\hat{V}_{j,\,\mu},\hat{V}_{j,\,\nu}^\dag
\hat{\rho}]\}.
\end{eqnarray}
here $J_{\mu,\nu}^{(j)}(\omega_{j,\nu})$ is the spectral density
of the jth reservoir,
\begin{eqnarray}\label{spectral density}
J_{\mu,\,\nu}^{(j)}(\omega_{j,\,\nu})=\int_0^\infty d\tau
\emph{e}^{i \omega_{j,\,\nu} \tau} G_{\alpha \beta}(\tau)
\end{eqnarray}
where $ G_{\alpha \beta}(\tau)$ is the environment
self-correlation function,
\begin{eqnarray}\label{G}
G_{\alpha \beta}(\tau)=Tr_{Bj}[\rho_{Bj}
\textit{\={f}}_{j,\,\nu}(\tau) \textit{\^f}_{j,\,\mu}]
\end{eqnarray}
and $ \textit{\={f}}_{j,\,\nu}(\tau)=\emph{e}^{-i
H_{Bj}\tau}\textit{\^f}_{j,\,\mu}^{\,\,\,\dag}\emph{e}^{i
H_{Bj}\tau}$. Spectral densities encapsulate the physical properties
of the environment and play an immensely important role in the
theoretical and experimental studies of the decoherence. In this
paper, we will consider the bosonic thermal bath as an infinite set
of harmonic oscillators  and apply a Weisskpof-Wignner-like
expression for spectral density such as
$J^{(j)}(\omega_{\mu})=\gamma_j(\omega_\mu) n_j(\omega_\mu)$ where
$n_j(\omega_\mu)=(\emph{e}^{\beta_j \omega_{\mu}}-1)^{-1}$ denotes
the thermal mean value of the number of excitations in the jth
reservoir at frequency $\omega_\mu$ and temperature
$T_j=\frac{1}{\beta_j}$  and $\gamma_j(\omega_\mu)$ is the coupling
strength of nanosystem and the jth reservoir. For simplicity we take
$\gamma_j(\omega_\mu)=\gamma_j$. Thus, the dissipators
$\pounds_j(\hat\rho)$ become

\begin{eqnarray}\label{dissipiators 2}
\pounds_j(\hat{\rho})&=&\sum_{\mu=1}^4 J^{(j)}(-\omega_{\mu})(2
\hat{V}_{j,\,\mu}\hat \rho
\hat{V}_{j,\,\mu}^{\,\,\,\,\dag}-\{\hat\rho,\hat{V}_{j,\,\mu}^{\,\,\,\,\dag}\hat{V}_{j,\,\mu}\}_+))\nonumber\\&+&
\sum_{\mu=1}^4 J^{(j)}(\omega_{\mu})(2
\hat{V}_{j,\,\mu}^{\,\,\,\,\dag}\hat \rho
\hat{V}_{j,\,\mu}-\{\hat\rho,\hat{V}_{j,\,\mu}\hat{V}_{j,\,\mu}^{\,\,\,\,\dag}\}_+))
\end{eqnarray}
with the transition frequencies
\begin{eqnarray}\label{frequencies}
\omega_1=\xi-\eta, \,\,\,\,\,\,\,\,\,\,\,\,\,\,\,\ \omega_4=-\omega_1,\nonumber\\
 \omega_2=\xi+\eta, \,\,\,\,\,\,\,\,\,\,\,\,\,\,\,\
 \omega_3=-\omega_2,
 \end{eqnarray}
 and the transition operators
\begin{eqnarray}\label{operators}
\hat V_{j,\,1}&=&a_{j,\,1} \outprod {\Psi^+}{\Sigma^+},\nonumber\\
\hat V_{j,\,2}&=&a_{j,\,2} \outprod {\Psi^+}{\Sigma^-},\\
\hat V_{j,\,3}&=&a_{j,\,3} \outprod {\Psi^-}{\Sigma^+},\nonumber\\
\hat V_{j,\,4}&=&a_{j,\,4} \outprod {\Psi^-}{\Sigma^-}, \nonumber\\
 \end{eqnarray}
 where
\begin{eqnarray}\label{coefficients}
\mid a_{j,\,1}\mid^2&=&\mid a_{j,\,4}\mid^2=\frac{1}{2 \xi
\eta}(\xi \eta
+J^2 \chi+(-1)^j B b),\nonumber\\
\mid a_{j,\,2}\mid^2&=&\mid a_{j,\,3}\mid^2=\frac{1}{2 \xi
\eta}(\xi \eta -J^2 \chi-(-1)^j B b),
 \end{eqnarray}
can be obtained from the spectrum of the nanosystem Hamiltonian.\\

The quantum master equation (\ref{master equation}) has an
important property, when the spectrum of $\hat H_s$ (see
eq.(\ref{spectrum})) is non-degenerate: In the energy basis,
$\{\ket{\varepsilon _i}\}_{i=1}^4$, the equations for diagonal
elements decouple from nondiagonal ones \cite{BP-book}.
Furthermore, nondiagonal elements are not coupled and the time
dependence of these elements has the simple form
\begin{eqnarray}\label{non diagonal}
\rho_{i,j}(t)= \rho_{i,j}(0) \emph e^{\alpha_{ij}t},
 \end{eqnarray}
where $\alpha_{i,j}\in\mathbb{C}$ are determined by the nanosystem
parameters. The equations for diagonal elements have the following
form
\begin{eqnarray}\label{diagonal}
\dot{R}(t)=B R(t),
 \end{eqnarray}
 where dot denotes the time derivative,
 $R(t)=(\rho_{11}(t), \rho_{22}(t), \rho_{33}(t), \rho_{44}(t))^T$
and B is the time independent $4\times4$ matrix

\begin{eqnarray}\label{B explicit form}
B  = \left( {\begin{array}{*{20}c}
    -(X_1^-+Y_2^-)&0 & X_1^+ & Y_2^+  \\
   0 & -(X_1^++Y_2^+)& Y_2^- & X_1^- \\
    X_1^-  &  Y_2^+ &  -(X_1^++Y_2^-) & 0 \\
    Y_2^-  & X_1^+ & 0 & -(X_1^-+Y_2^+)    \\
\end{array}} \right),
\end{eqnarray}
where

\begin{eqnarray}\label{XY1}
X_\mu^\pm=2 \sum_{j=1,2}J^{(j)}(\mp\omega_\mu) \mid a_{j,\,1}\mid^2,\nonumber\\
Y_\mu^\pm=2 \sum_{j=1,2}J^{(j)}(\mp\omega_\mu) \mid
a_{j,\,2}\mid^2.
 \end{eqnarray}
 The analytical solution of the equation (\ref{diagonal}) in the
 energy basis is given by
\begin{eqnarray}\label{R t}
R(t)=M(t) R(0)
\end{eqnarray}
where $M(t)=[m_{i\,j}]_{4\times4}$, and the elements $m_{ij}$ are
defined by

\begin{eqnarray}\label{XY2}
m_{1\,1}&=&\frac{1}{X_1 Y_2}(X_1^++X_1^- \emph e^{-tX_1})(Y_2^++Y_2^- \emph e^{-t Y_2}),\nonumber\\
m_{1\,2}&=&\frac{1}{X_1 Y_2}(1- \emph e^{-t X_1})(1- \emph e^{-t Y_2})X_1^+Y_2^+,\nonumber\\
m_{1\,3}&=&\frac{1}{X_1 Y_2}(1- \emph e^{-t X_1})X_1^+(Y_2^++Y_2^-\emph e^{-t Y_2}),\nonumber\\
m_{1\,4}&=&\frac{1}{X_1 Y_2}(X_1^++X_1^- \emph e^{-tX_1})(1- \emph e^{-t Y_2})Y_2^-,\nonumber\\
m_{2\,1}&=&\frac{1}{X_1 Y_2}(1- \emph e^{-t X_1})(1- \emph e^{-t Y_2})X_1^-Y_2^-,\nonumber\\
m_{2\,2}&=&\frac{1}{X_1 Y_2}(X_1^-+X_1^+ \emph e^{-tX_1})(Y_2^-+Y_2^+ \emph e^{-t Y_2}),\nonumber\\
m_{2\,3}&=&\frac{1}{X_1 Y_2}(X_1^-+X_1^+ \emph e^{-tX_1})(1- \emph e^{-t Y_2})Y_2^-,\nonumber\\
m_{2\,4}&=&\frac{1}{X_1 Y_2}(1- \emph e^{-t X_1})X_1^-(Y_2^-+Y_2^+ \emph e^{-t Y_2}),\nonumber\\
m_{3\,1}&=&\frac{1}{X_1 Y_2}(1- \emph e^{-t X_1})X_1^-(Y_2^++Y_2^-\emph e^{-t Y_2}),\nonumber\\
m_{3\,2}&=&\frac{1}{X_1 Y_2}(X_1^-+X_1^+ \emph e^{-tX_1})(1- \emph e^{-t Y_2})Y_2^+,\nonumber\\
m_{3\,3}&=&\frac{1}{X_1 Y_2}(X_1^-+X_1^+ \emph e^{-tX_1})(Y_2^++Y_2^- \emph e^{-t Y_2}),\nonumber\\
m_{3\,4}&=&\frac{1}{X_1 Y_2}(1- \emph e^{-t X_1})(1- \emph e^{-t Y_2})X_1^-Y_2^+,\nonumber\\
m_{1\,4}&=&\frac{1}{X_1 Y_2}(X_1^++X_1^- \emph e^{-tX_1})(1- \emph e^{-t Y_2})Y_2^-,\nonumber\\
m_{4\,2}&=&\frac{1}{X_1 Y_2}(1- \emph e^{-t X_1})X_1^+(Y_2^-+Y_2^+ \emph e^{-t Y_2}),\nonumber\\
m_{4\,3}&=&\frac{1}{X_1 Y_2}(1- \emph e^{-t X_1})(1- \emph e^{-t Y_2})X_1^+Y_2^-,\nonumber\\
m_{4\,4}&=&\frac{1}{X_1 Y_2}(X_1^++X_1^- \emph e^{-tX_1})(Y_2^-+Y_2^+ \emph e^{-t Y_2}),\nonumber\\
\end{eqnarray}
here we have defined: $ X_\mu=X_\mu^++X_\mu^-$ and
$Y_\mu=Y_\mu^++Y_\mu^-$.\\
There is a singular point $\xi=\eta$, for which the spectrum
(\ref{spectrum}) becomes degenerate and the above solution is not
valid. The state of the system is not well defined at this critical
point. This critical point assigns a critical value for the
parameters of the nanosystem such as critical magnetic field
($B_c$), critical parameter of inhomogeneity of magnetic field
($b_c$), critical spin-orbit interaction parameter ($D_c$) and etc.
The behavior of the entanglement of the system changes abruptly when
the parameters cross their critical values (see section III).
\\
In the following we will examine a class of bipartite density
matrices having the following standard form as the initial state of
the system \footnote{The subscript "s" denotes the standard
computational basis $\{\ket{00},\ket{01},\ket{10},\ket{11}\}$.}
\begin{eqnarray}\label{initial dm}
\hat\rho_s(0)  = \left( {\begin{array}{*{20}c}
   {\mu _ +  } & 0 & 0 & \nu   \\
   0 & {w_1 } & z & 0  \\
   0 & {z^* } & {w_2 } & 0  \\
   \nu  & 0 & 0 & {\mu _ -  }  \\
\end{array}} \right).
\end{eqnarray}
These kind of density matrix are  called X states class and arises
naturally in a wide variety of physical situations. This class
contains some important subsets like the pure Bell states, the
states which can be expressed as a mixture of Bell states, Werner
states and so on. If the initial state, $\hat\rho_s(0)$, belongs to
the set of X states (\ref{initial dm}) then Eq (\ref{spectrum})
guarantees that $\hat\rho_s(t)$ given by Eqs. (\ref{R t} and
\ref{non diagonal}) also belongs to the same set. Therefore, the
only
 non-vanishing off diagonal components of the density matrix in
 the energy basis are
\begin{eqnarray}\label{R t}
\rho_{12}(t)&=&\rho_{12}(0) \emph e ^{-2 i \xi t-t(
X_1+Y_2)/2}, \,\,\,\,\,\,\,\,\,\,\,\,\,\,\,\rho_{21}(t)=\rho_{12}(t)^*,\nonumber\\
\rho_{34}(t)&=&\rho_{34}(0) \emph e ^{-2 i \eta t-t( X_1+Y_2)/2},
\,\,\,\,\,\,\,\,\,\,\,\,\,\,\,\rho_{43}(t)=\rho_{34}(t)^*.
\end{eqnarray}
Knowing the density matrix we can calculate the concurrence
$C(\rho(t))=\max\{0,2\lambda_{max}(t)-\sum_{i=1}^{4}\,\lambda_i(t)\}$
where,
\begin{eqnarray}\label{lambda t}
\lambda_{1,\,2}(t)&=&\mid \sqrt{\rho_{s11}(t) \rho_{s44}(t)}\pm
\mid
\rho_{s14}(t)\mid\,\,\mid, \nonumber\\
\nonumber\\
 \lambda_{3,\,4}(t)&=&\mid\sqrt{\rho_{s22}(t) \rho_{s33}(t)}\pm \mid
\rho_{s23}(t)\mid\,\,\mid.
\end{eqnarray}
 Unfortunately, the $\lambda_i(t)$s depend on the parameters involved.
  This prevents us from writing an analytical expression for concurrence.
  But it is possible to evaluate concurrence, numerically for a given set of the
parameters. The results are shown in Figs. 1-7. Figs. 1-3 depict the
dynamical behavior of concurrence versus parameters of the
nanosystem and reservoirs and Figs. 4-7 illustrate the asymptotical
behavior of the concurrence versus parameters of the nanosystem and
reservoirs. Without loss of generality we can assume that $J>0$,
since the above formula are invariant under substitution
$J\longrightarrow -J$. This means that
the dynamical behavior of the FM chain is the same as the AFM chain.\\

\subsection{Asymptotic case}

For a class of states, $\hat\rho_{st}$, the dissipative and
decoherence mechanisms (second and third terms in the master
equation (\ref{master equation})) compensate the unitary dynamics
which is governed by nanosystem Hamiltonian (first term in master
equation (\ref{master equation})) i.e. $i
[\hat{H}_S,\hat\rho_{st}]=\pounds_1(\hat\rho_{st})+\pounds_2(\hat\rho_{st})$
or $\frac{d }{dt}\hat \rho_{st}=0$. These states are called
stationary states because they are constant in time. If there exist
such a stationary state solution for the master equation, the system
tends to it asymptotically in large time limit i.e.
$\lim_{t\rightarrow\infty} \hat \rho(t)\longrightarrow \hat
\rho_{asym.}=\hat\rho_{st}$. For the present system (thermal
reservoirs and interacting nanosystem),
 nondiagonal elements (\ref{non diagonal}) vanish asymptotically at large time limit and hence
  $\hat\rho(t)$ converges to a diagonal density matrix (in the energy basis) with elements
   not depending on the initial conditions:
\begin{eqnarray}\label{rho asym1 }
\hat \rho_{asym.}=\frac{1}{X_1 Y_2} \,\,\,\textrm{diagonal}(X_1^+
Y_2^+,X_1^- Y_2^-,X_1^- Y_2^+,X_1^+ Y_2^-).
\end{eqnarray}
The asymptotic concurrence is given by
$C(\rho_{asym.})=C^\infty=\max\{0,2\lambda_{max}-\sum_{i=1}^{4}\,\lambda_i\}$
with
\begin{eqnarray}\label{lambda asym}
\lambda_{1,\,2}&=&\mid \sqrt{\rho_{s11}^{asym.}
\rho_{s44}^{asym.}}\pm \mid
\rho_{s14}^{asym.}\mid\,\,\mid, \nonumber\\
\nonumber\\
 \lambda_{3,\,4}&=&\mid\sqrt{\rho_{s22}^{asym.} \rho_{s33}^{asym.}}\pm \mid
\rho_{s23}^{asym.}\mid\,\,\mid,
\end{eqnarray}
where
\begin{eqnarray}\label{rho asym2}
\rho_{s11}^{asymp.}&=&\frac{1}{2 \eta X_1 Y_2}((\eta +B)X_1^-
Y_2^++(\eta -B)X_1^+ Y_2^-), \nonumber\\
\rho_{s14}^{asymp.}&=&\frac{J \chi}{2 \eta X_1 Y_2}(X_1^-
Y_2^+-X_1^+ Y_2^-)=\rho_{s41}^{asymp.}, \nonumber\\
\rho_{s22}^{asymp.}&=&\frac{1}{2 \xi X_1 Y_2}((\xi +b)X_1^+
Y_2^++(\xi -b)X_1^- Y_2^-), \nonumber\\
\rho_{s23}^{asymp.}&=&\frac{J(1+i D)}{2 \xi X_1 Y_2}(X_1^+
Y_2^+-X_1^- Y_2^-)=(\rho_{s32}^{asymp.})^*, \nonumber\\
\rho_{s33}^{asymp.}&=&\frac{1}{2 \xi X_1 Y_2}((\xi -b)X_1^+
Y_2^++(\xi +b)X_1^- Y_2^-), \nonumber\\
\rho_{s44}^{asymp.}&=&\frac{1}{2 \eta X_1 Y_2}((\eta -B)X_1^-
Y_2^++(\eta +B)X_1^+ Y_2^-).
\end{eqnarray}
There is an interesting limiting case for which the coupled QDs are
in contact with two independent reservoirs at identical temperatures
($\beta_1=\beta_2=\beta$). In this case, it is easy to show that

\begin{eqnarray}\label{XY 3}
\frac{X_1^+}{X_1}&=&\frac{e^{\beta \omega_1}}{e^{\beta
\omega_1}+1},
\,\,\,\,\,\,\,\,\,\,\,\,\,\,\,\,\,\,\frac{X_1^-}{X_1}=\frac{1}{e^{\beta
\omega_1}+1} \nonumber\\
\frac{Y_2^+}{Y_2}&=&\frac{e^{\beta \omega_2}}{e^{\beta
\omega_2}+1},
\,\,\,\,\,\,\,\,\,\,\,\,\,\,\,\,\,\,\frac{Y_2^-}{Y_2}=\frac{1}{e^{\beta
\omega_2}+1}.\nonumber
\end{eqnarray}
By substituting these relations into the Eq. (\ref{rho asym1 }), the
reduced density matrix $\hat\rho_{asym.}$ takes the thermodynamical
canonical form for a system described by the Hamiltonian $\hat H_S$
at temperature $T=\beta^{-1}$, as expected. This means that
\begin{eqnarray}\label{rho asym eq}
\hat\rho_{asym.}(\Delta T=0)\equiv\hat\rho_T=\frac{e^{-\beta
H_S}}{Z},
\end{eqnarray}
where $Z=Tr(e^{-\beta H_S})$ is the partition function. Thermal
entanglement properties of such systems have been studied
substantially, in our previous work \cite{Hamid}. Thus, for the
special case $\Delta T=0$, the results coincide with the results
of Ref.\cite{Hamid}.

\section{Results}

 The non-equilibrium thermal concurrence as a function of time for
three values of temperature difference ($\Delta T=T_1-T_2$) and for
a fixed value of mean temperature ($T_M=\frac{T_1+T_2}{2}$) are
plotted in Fig. 1, for  the  case of "direct geometry" of connection
($b \Delta T>0$). The presence of temperature difference has not
effective influence on the dynamics of entanglement at early times
of evolution, but changes effectively the behavior of the asymptotic
entanglement (which is more evident form Figs. 4-7). Fig. 2 depicts
the variation of entanglement dynamics for some values of $T_M$ and
for a fixed $\Delta T$ in the case of "direct geometry" of
connection ($b \Delta T>0$). By Increasing $T_M$, thermal
fluctuations suppress quantum fluctuations  and hence decrease
coherent oscillations at early
times of evolution and also decrease asymptotic entanglement. \\
Perhaps surprisingly, the decoherence due to environmental
interaction does not prevent the creation of a steady state level of
entanglement, regardless of the initial state of the system. This is
demonstrated in figure 3 which shows the time evolution of the
non-equilibrium thermal concurrence for a given set of parameters
and for four different initial states: (i) a maximally entangled
state, the Bell state, $\ket{\psi(0)}=\frac{1}{\sqrt 2}(\ket
{01}+\ket{10})$  (ii) a separable state, $\ket{\psi(0)}=\ket{01}$
 (iii) a mixed state, defined as an equal mixture of a Bell state
and a product state, e.g.
$\rho_s(0)=\frac{1}{4}(\ket{00}+\ket{11})(\bra{00}+\bra{11})+\frac{1}{2}
\proj{01}{01}$ and finally (iv) an unpolarized state,
$\rho_s(0)=\frac{1}{4}I$. Despite the presence of decoherence (due
to interaction with environment) the results of figure 3 show that
the concurrence reaches the same steady state value, $C^\infty$
(after some oscillatory behavior) for a given set of parameters,
regardless of initial state of the system. Clearly the Heisenberg
interaction in the Eq. (\ref{Hamiltonian 2}) serves to maintain an
entangled asymptotic state despite the presence of decoherence. At
early times of evolution the amount of $D$ determines the
frequency of oscillations. Each plot in Fig. 3 contains two cases
a) $D<D_c$, in this case asymptotic value of entanglement
decreases with $D$ and b) $D>D_c$, in this case
asymptotic value of entanglement increases as $D$ increases.\\
In Fig. 4, the asymptotic non-equilibrium thermal concurrence is
plotted versus $T_M$ and $D$ for different values of $\Delta T$. For
the case of identical temperatures ($\Delta T=0$) the results are
the same as ref. \cite{Hamid}, as mentioned in the previous section.
This figure shows that there is a critical mean
temperature($T_M^{cr.}$) over which entanglement vanishes (ESD
phenomenon). The size of $T_M^{cr.}$ and the amount of entanglement
can be improved by increasing $D$. For Small values of $D$
($D<D_c$), an increase in $\Delta T$ increases the size of
$T_M^{cr.}$ i.e entanglement can exists in higher mean temperatures
due to existence of temperature difference of
reservoirs (environment induced entanglement). \\
The variation of the asymptotic non-equilibrium thermal concurrence
as a function of $\Delta T$ and $D$ for fixed values of $T_M$ and
$b$ is illustrated in Fig. 5. The behavior of concurrence is
dependent on the geometry of connection. For the case of "direct
geometry" ($b \Delta T>0$) and for $D<D_c$, no entanglement is
observed but for the same geometry and for $D>D_c$ the amount of
entanglement is nonzero and it increases as $D$ or $\Delta T$
increases. For the case of "indirect geometry" ($b \Delta T<0$),
there is a nonzero entanglement for all values of $D$. The amount of
entanglement is an increasing function of $D$ and increases with
$\Delta T$ for the values of $\Delta T\leq\frac{T_M}{2}$ and
decreases with $\Delta T$ for the values of $\frac{T_M}{2}<\Delta
T \leq T_M$.\\
Figs. 6 and 7 show the behavior of the asymptotic non-equilibrium
thermal concurrence versus $T_M$ and $\Delta T$ for different values
of $D$ and for the symmetric ($b=0$) and nonsymmetric ($b\neq0$)
cases, respectively. Both figures reveal that, increasing $D$ cause
to the appearance of entanglement in the larger region of
$T_M-\Delta T$ plane. The departure between symmetric (Fig. 6) and
nonsymmetric (Fig. 7) case is more obvious for $D<D_c$. For
nonsymmetric case, Fig.7 shows that the "indirect geometry" of
connection is more suitable for creating entanglement when $D<D_c$.
In both symmetric and nonsymmetric cases, maximum entanglement
($C=1$) can be achieved in the case of identical temperatures
($\Delta T=0$), zero mean temperature ($T_M=0$) (i.e when both of
reservoirs are in the ground state ($T_1=T_2=0$)) and large values
of $D$ (it is in agreement with the results of \cite {Hamid}).

\section{Discussion}

The Dynamics of non-equilibrium thermal entanglement of an open
two-qubit nanosystem is investigated. The inter-qubit interaction is
considered as the Heisenberg interaction in the presence of
inhomogeneous magnetic field and spin-orbit interaction (arises from
the Dzyaloshinski- Moriya (DM) anisotropic antisymmetric
interaction). Each qubit interacts with separate thermal reservoir
(bosonic bath) which is held in its own temperature. The effects of
the parameters of the model, including the parameters of the
nanosystem (especially, the parameter of the spin orbit interaction
(D)) and environmental parameters (particularly, mean temperature
$T_M$ and temperature difference $\Delta T$)), on the
non-equilibrium thermal entanglement dynamics of the nanosystem is
investigated, by solving the quantum Markov-Born master equation of
the nanosystem. An analytical solution of the master equation is
derived and then entanglement dynamics and asymptotic entanglement
of the nanosystem versus the parameters of the model is studied.
Resolving the entanglement dynamics allowed us to distinguish
between entanglement induced by the interaction and by the
environment. The results show that, decoherence induced by thermal
bathes are competing with inter-qubit interaction terms to create a
steady state level of entanglement, as measured by the concurrence.
The size of this steady state (asymptotic) entanglement and the
dynamical behavior of entanglement are dependent on the parameters
of the model and also depend on the geometry of connection.
Increasing temperature difference $\Delta T$, and mean temperature
$T_M$, kill the asymptotic entanglement. Indeed, thermal
fluctuations suppress the quantum fluctuations (i.e. all quantum
effects such as the entanglement and local coherence), and hence the
entanglement of the system dies at a critical temperature
$T_M^{cr.}$ (ESD). We have shown that, the size of $T_M^{cr.}$ and
the amount of entanglement can be enhanced by choosing a suitable
value of spin-orbit interaction parameter $D$. The maximum
entanglement ($C=1$) can be achieved for the case of large values of
$D$ and zero temperature reservoirs($T_1=T_2=0$).
 For physical realization of the model we address a two
coupled quantum dots which are interacting with two independent
thermal bathes. Furthermore, we find that the indirect geometry of
connection is more suitable for creating and maintaining the
entanglement. In this case, the heat current between two QDs is
substantially decreased but the concurrence can enhances by
temperature difference $\Delta T$ \cite{QR}. For the special case
$\Delta T=0$, our results confirm the results of \cite{Hamid}. The
results of \cite{QR, Si} are also obtained for the special case of
$D=\chi=0$ and by considering $b=\epsilon_1-\epsilon_2$. The results
can provide useful recipes for realistic quantum information
processing in noisy and non-equilibrium environments.

\newpage

\begin{figure}
\epsfxsize=11cm \ \centerline{\hspace{0cm} \epsfbox{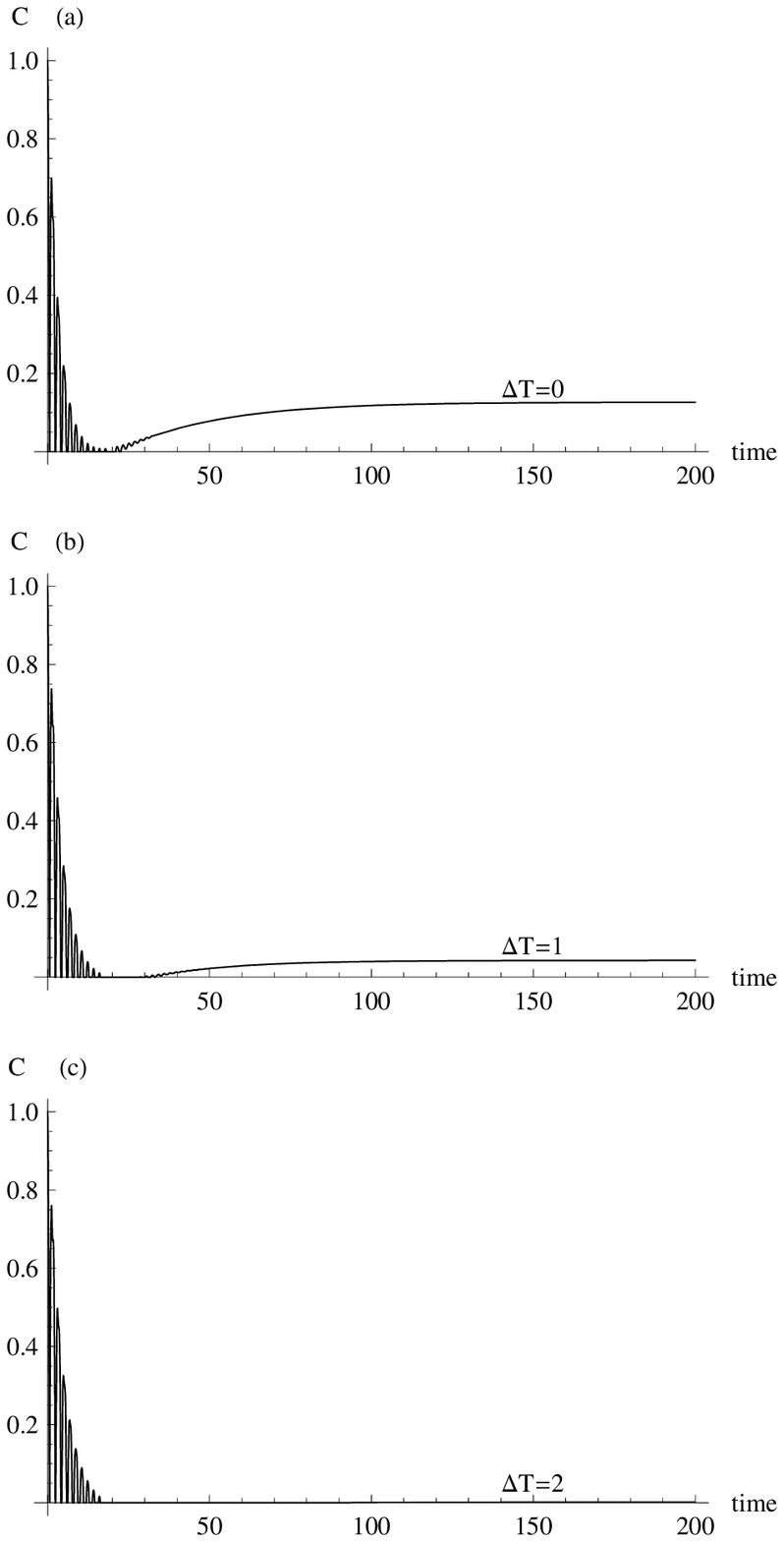}}
\ \caption{ Dynamics of non-equilibrium concurrence for the
initial reduced density matrix
$\rho_s(0)=\frac{1}{\sqrt{2}}(\proj{01}{01}+\proj{10}{10})$. The
parameters of the model are chosen to be $\gamma_1=\gamma_2=0.02$,
$J=1$, $\chi=0.9$, $B=2$, $b=1$, $T_M=1.5$ and for different
values of temperature difference $\Delta T$: (a) $\Delta T=0$ (b)
$\Delta T=1$ (c) $\Delta T=2$. All parameters are
dimensionless.}\label{f1}
\end{figure}

\begin{figure}
\epsfxsize=11cm \ \centerline{\hspace{0cm}\epsfbox{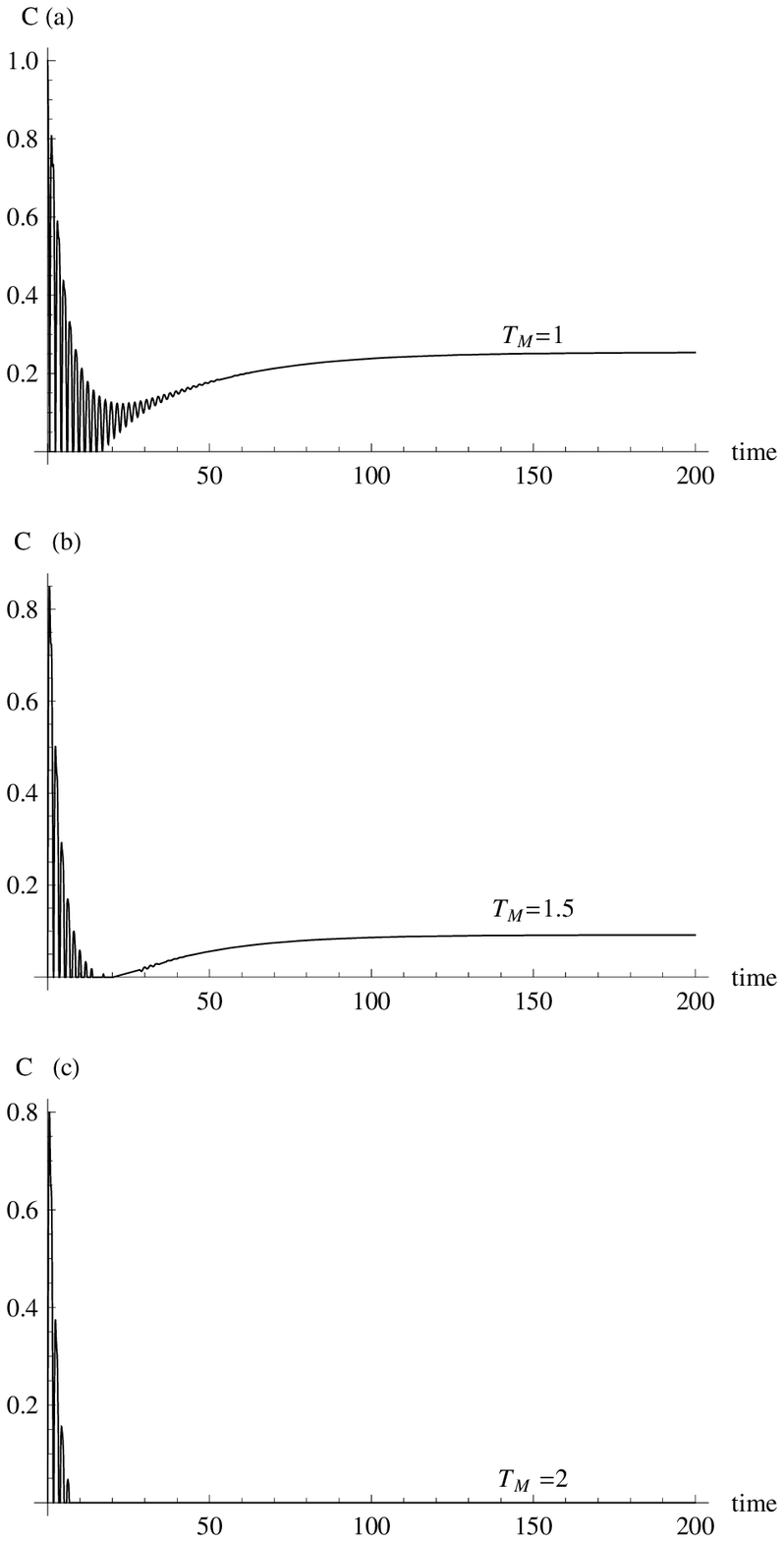}} \
\caption{Dynamics of non-equilibrium concurrence for the initial
reduced density matrix
$\rho_s(0)=\frac{1}{\sqrt{2}}(\proj{01}{01}+\proj{10}{10})$. The
parameters of the model are chosen to be $\gamma_1=\gamma_2=0.02$,
$J=1$, $\chi=0.9$, $B=2$, $b=1$, $\Delta T=0.5$ and for different
values of mean temperature $T_M$: (a)
 $T_M=1$ (b)  $T_M=1.5$ (c) $T_M=2$. All parameters are
dimensionless.}\label{f2}
\end{figure}

\begin{figure}
\epsfxsize=16cm \ \centerline{\hspace{0cm}\epsfbox{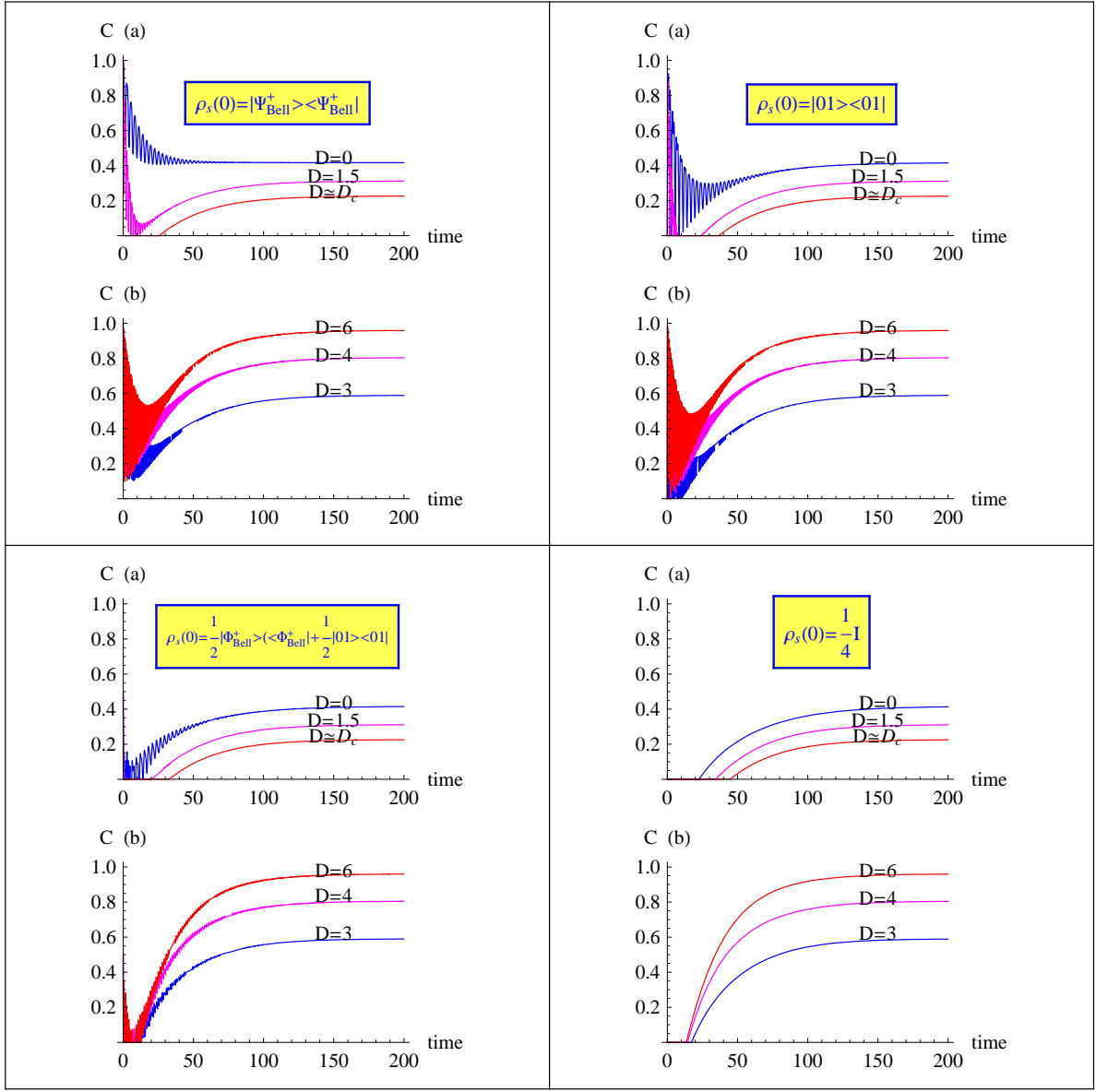}} \
\caption{(Color online) Dynamics of non-equilibrium concurrence
for different values of $D$ and different initial state. The
parameters of the model are chosen to be $\gamma_1=\gamma_2=0.02$,
$J=1$, $\chi=0.9$, $B=2$, $b=0.5$,  $T_M=1$ and $\Delta T=0.5$.
Each plot contains two graphs for (a) $D<D_c$ (b) $D>D_c$
($D_c\simeq1.8868$). All parameters are dimensionless. }\label{f3}
\end{figure}

\begin{figure}
\epsfxsize=12cm \ \centerline{\hspace{0cm}\epsfbox{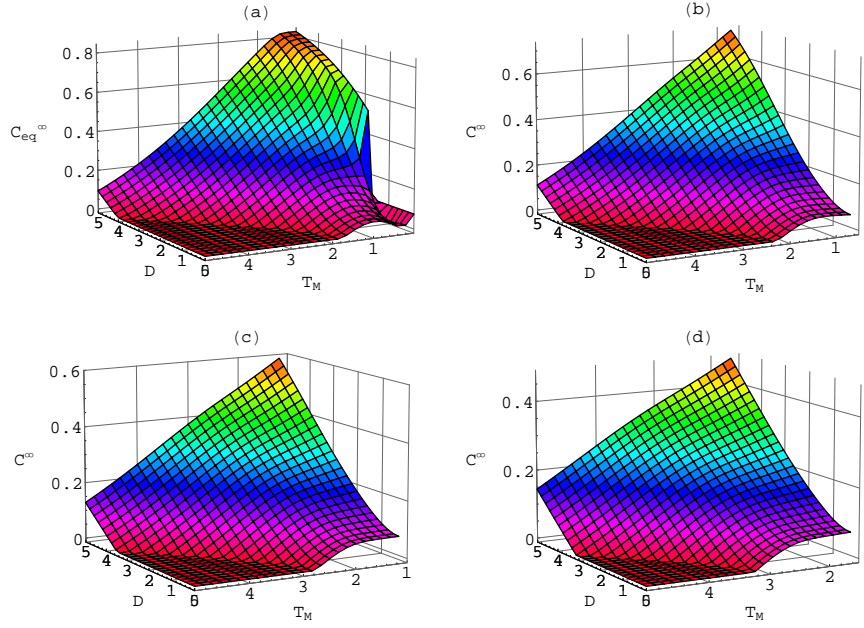}} \
\caption{(Color online) Asymptotic entanglement vs. $T_M$ and D,
The parameters of the model are chosen to be
$\gamma_1=\gamma_2=0.02$, $B=4$, $J=1$, $\chi=0.3$ and $b=-3.5$
for (a) $\Delta T=0$ (b) $\Delta T=1$ (c) $\Delta T=2$ (d) $\Delta
T=3$ ($D_c\simeq1.68523$). All parameters are dimensionless.
}\label{f4}
\end{figure}

\begin{figure}
\epsfxsize=13.5cm \ \centerline{\hspace{0cm}\epsfbox{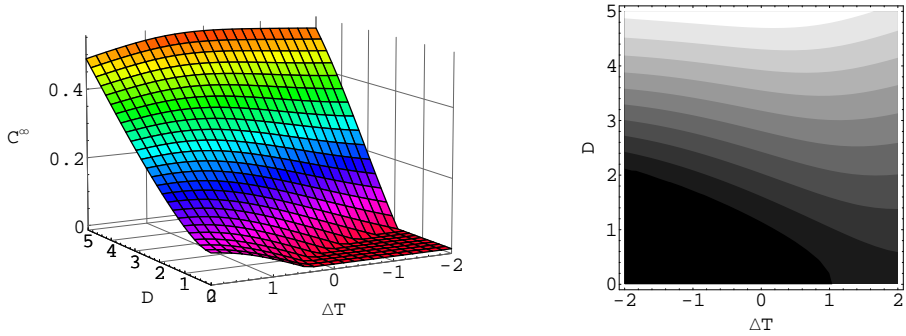}}
\ \caption{(Color online) Asymptotic entanglement vs. $\Delta T$
and D. The parameters of the model are chosen to be
$\gamma_1=\gamma_2=0.02$, $B=4$, $J=1$, $\chi=0.3$ and $b=-3.5$
and $T_M=2$ ($D_c\simeq1.68523$). All parameters are
dimensionless.}\label{f5}
\end{figure}

\begin{figure}
\epsfxsize=13.5cm \ \centerline{\hspace{0cm}\epsfbox{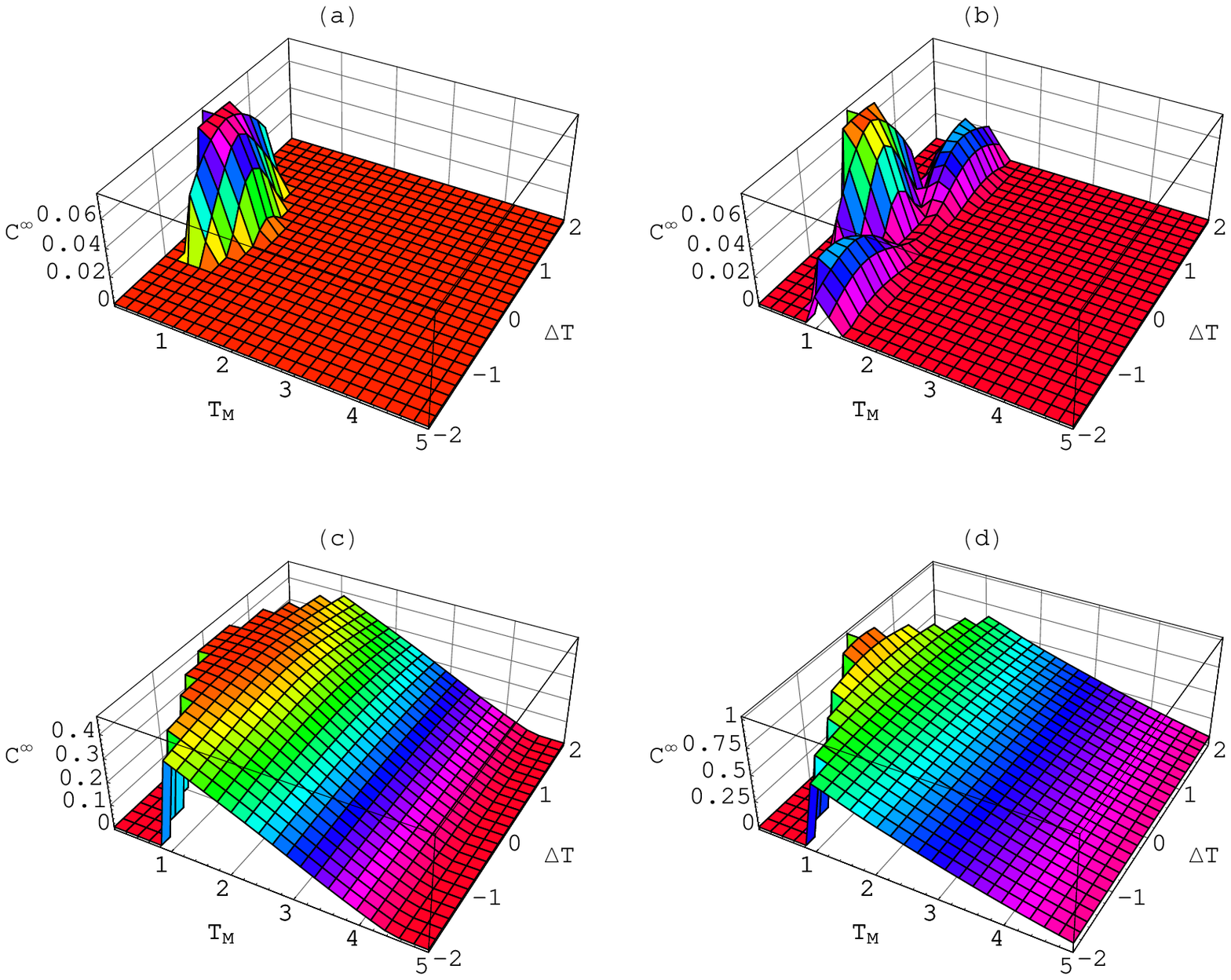}}
\ \caption{(Color online) Asymptotic entanglement vs. $T_M$ and
$\Delta T$. The parameters of the model are chosen to be
$\gamma_1=\gamma_2=0.02$, $B=4$, $J=1$, $\chi=0.3$ and $b=0$ for
(a) $D=0$ (b) $D=1$ (c) $D \approx D_c$ (d) $D=5$
($D_c\simeq3.88458$). All parameters are dimensionless.}\label{f6}
\end{figure}

\begin{figure}
\epsfxsize=13.5cm \ \centerline{\hspace{0cm}\epsfbox{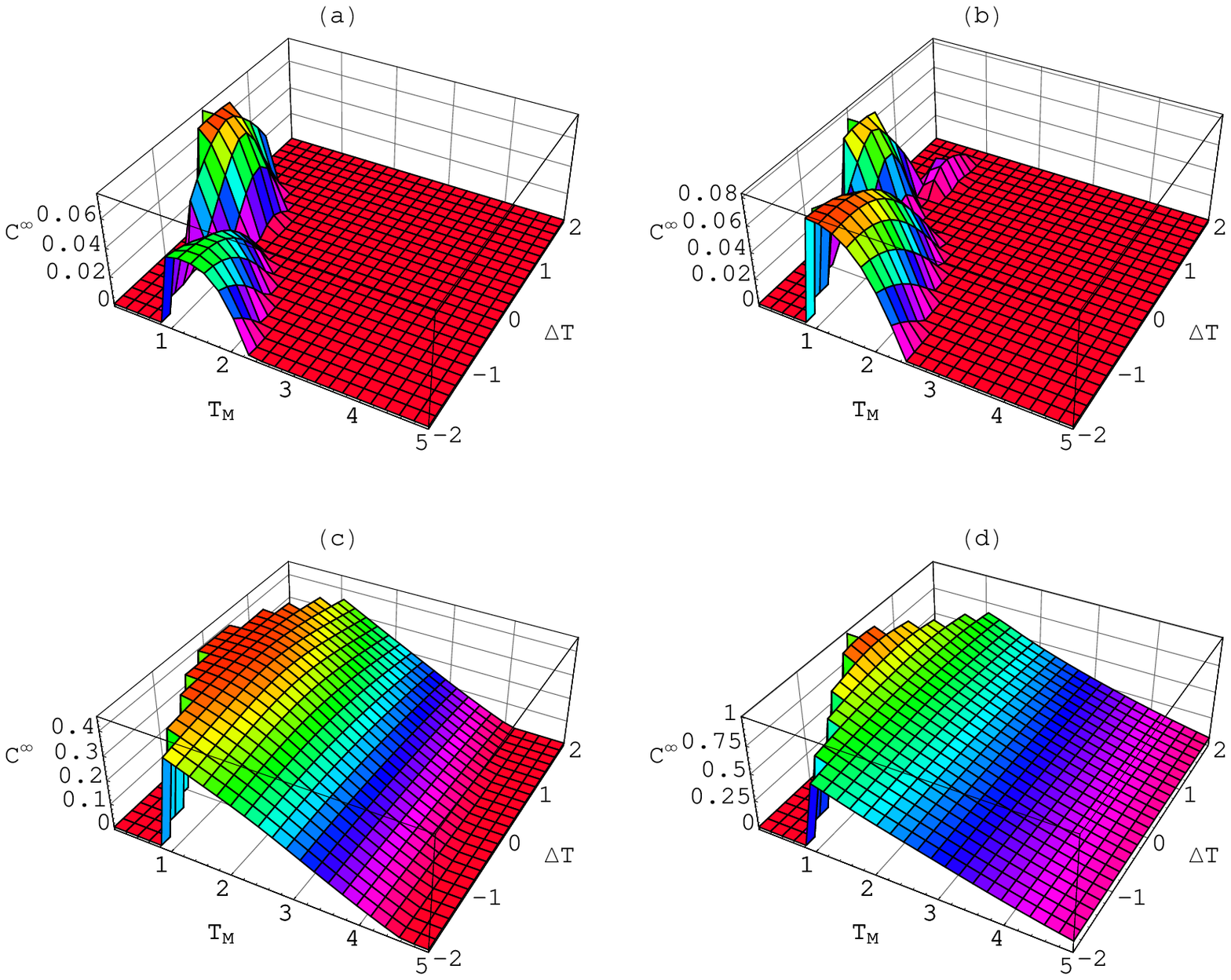}}
\ \caption{(Color online) Asymptotic entanglement vs. $T_M$ and
$\Delta T$. The parameters of the model are chosen to be
$\gamma_1=\gamma_2=0.02$, $B=4$, $J=1$, $\chi=0.3$ and $b=1$ for
(a) $D=0$ (b) $D=1$ (c) $D \approx D_c$ (d) $D=5$
($D_c\simeq3.75366$). All parameters are dimensionless.}\label{f7}
\end{figure}
\end{document}